\documentclass[runningheads]{llncs}
\usepackage{graphicx}
\usepackage{todonotes}
\usepackage{lipsum}

\begin{document}
%

\title{Towards a decentralized data privacy protocol for self-sovereignty in the digital world\thanks{This work has been funded by the German Federal Ministry of Education and Research (BMBF) (grant numbers 16KIS1507 and 16KIS1510).}}


%
\titlerunning{Towards a decentralized data privacy protocol}
%
\author{Rodrigo Falc\~{a}o\inst{1}\orcidID{0000-0003-1222-0046} \and
Arghavan Hosseinzadeh\inst{1}\orcidID{0009-0001-8699-9972}}
\authorrunning{R. Falc\~{a}o and A. Hosseinzadeh}
%
\institute{Fraunhofer IESE, Kaiserslautern 67663, Germany  \\
\email{\{rodrigo.falcao, arghavan.hosseinzadeh\}@iese.fraunhofer.de}\\}
\maketitle              
\begin{abstract}
A typical user interacts with many digital services nowadays, providing these services with their data. As of now, the management of privacy preferences is service-centric: Users must manage their privacy preferences according to the rules of each service provider, meaning that every provider offers its unique mechanisms for users to control their privacy settings. However, managing privacy preferences holistically (i.e., across multiple digital services) is just impractical. In this vision paper, we propose a paradigm shift towards an enriched user-centric approach for cross-service privacy preferences management: the realization of a decentralized data privacy protocol.

\end{abstract}

\section{Motivation}

Not long ago, users navigated through Internet-based services (e.g., websites) leaving behind an unnoticed trail of personal data, browsing habits, and online interactions. Tracking technologies such as cookies, the development of data analysis techniques, and the rise of tech giants such as Google and Facebook (whose online presence became pervasive) have given rise to a massive concentration of data, which has supported targeted advertising for diverse purposes, sometimes raising ethical concerns \cite{daoud2023examining}. These concerns led to the emergence of several data protection initiatives around the globe, resulting in regulations such as the GDPR \cite{gdpr2016} and the Data Governance Act \cite{eu2022datagovernance}, which aim to facilitate data exchange while increasing trust and ensuring data protection. One of the most tangible consequences a user experiences occurs when they visit a website using cookie technology: On first access, they are asked to accept or reject cookies.

While users have been given some sort of freedom to choose which types of data are used by different services and for what purposes, and have also been entitled to change their preferences and revoke permissions for data usage, it is still hard for them maintain sovereignty over their choices. Gradually, their data and their choices regarding data usage are spread across the dozens, hundreds, or even thousands of digital services they interact with. In the long run, it is easy to lose track of where data is stored and used, by whom, and for what purpose. Taking back control of the usage of their data becomes a near-impossible endeavor. Even if each digital service were shipped with a comprehensive and easy-to-use privacy preferences management tool, users would still have to remember all the services that use their data, visit them, and, one after another, review their privacy preferences if they want to. In other words, from the end user's perspective, privacy preferences management solutions are distributed across virtually all digital services they interact with. For this very reason, they cannot be effective. To make matters worse, there is a lack of standardization in the field. 

In this vision paper, we propose a paradigm shift in how we approach the issue of privacy preferences management. The current paradigm focuses on the interaction between a user and a digital service. Usable privacy management tools are a necessary but insufficient solution. We envision a future where a fully user-centric approach takes a holistic view of all digital services, allowing users to manage their preferences in one place, without any particular service provider monopolizing this space. This is where the decentralized data privacy protocol comes into play. It can be implemented by any party and offers benefits to both end users and service providers. It also opens up opportunities for the development of new privacy-enhancing technologies on top of it. The remainder of this paper is structured as follows: In Section~\ref{asp24:sec:current}, we review recent related work; Section~\ref{asp24:sec:protocol} outlines the protocol concept; Section~\ref{asp24:sec:opportunities} explores the benefits and opportunities; and Section~\ref{asp24:sec:conclusion} concludes the paper, outlining our next steps.


\section{Related work}\label{asp24:sec:current}

The most recent advancements in research with respect to preserving user privacy focus on privacy policy languages and user-friendly settings for privacy preferences. Gharib \cite{gharib2022toward} argues that most data subjects blindly accept the notices, not because they do not value their privacy, but because most privacy policies and terms of services are long, complex, and hard to understand, so the author introduces a model for informed consent. The model involves a Matching Component that compares the privacy preferences of the user that is included within their Personal Privacy Profile (PPPo) and the privacy policies that are published by the service providers and automates the process of giving consent. A dynamic contextual notice is provided to the user when the user preferences and the policies do not match. Accordingly, the user can make an informed decision concerning the consent request. Gharib also proposes an ontology that can be used for realizing preferences and policies. However, where to store the PPPos seems to be out of the scope of this work.

Dehling et al. \cite{dehling2024konzepte} introduce Privacy Cockpits, which serve as central dashboards for users to navigate and manage their personal data. This solution aims to ease the enforcement of regulations such as the GDPR, although the focus of their work is on protecting the data used across various services within specific digital ecosystems.

In 2020, the European Data Protection Supervisor introduced the Personal Information Management Systems (PIMS) concept \cite{edps2020techdispatch}. The PIMS concept offers a new approach in which individuals are the ``holders'' of their own personal information. PIMS aims to empower users to take charge of their digital identity and the use of their personal information across various services and platforms. In recent years, several initiatives and projects have claimed PIMS features. Among these, the Solid protocol \cite{sambra2016solid} stands out. It proposes a set of conventions and tools for building a decentralized platform for social Web applications. In order to address the challenge of obtaining consent for processing personal data, Florea and Esteves \cite{florea2023automated} introduced a policy layer into the Solid ecosystem. They integrated the usage of the ODRL Policy Language \cite{w3codrl2}, the ODRL profile for Access Control (OAC) \cite{odrlAccessControlProfile}, and the Data Privacy Vocabulary (DPV) \cite{w3cdpv} to allow Solid users to express their privacy preferences. By integrating such a policy layer into the Solid ecosystem, the matching process regarding users' preferences and requests for data can be automated. However, the focus of the Solid protocol remains on data management mechanisms that allow users to store their data such as contacts and photos in Personal Online Datastores (PODs) and control access of applications to this data. We therefore still see an ongoing need for a protocol specifically designed for privacy preferences management.





\section{A paradigm shift}\label{asp24:sec:protocol}


We propose a paradigm shift by adjusting the context of the problem. Users need adequate means to exercise their data sovereignty. Tools can implement diverse models to provide such means; however, regardless of whether we consider the implementation of the ``notice and consent model'', the ``informed consent model'', or ``data protection cockpits'', to name but a few strategies, solutions address the data sovereignty challenge in the context of the interaction between the user and \textit{a given digital service (or ecosystem)}. Consider now the exercise of data sovereignty in the context of not only one digital service, but \textit{all of them}. If we have this goal in mind, current strategies -- though valuable and necessary -- fall short. It is not enough that \textit{each digital service} provides its own means for users to manage their preferences; instead, \textit{each individual} should be able to manage their data preferences \textit{across all digital services that use their data}.
A new centralized digital service could fill this gap; however, to be effective, it would require the providers of digital services to adhere to it, and users would still have to rely on a centralized service offered by yet another service provider.

\begin{figure}[t]
    \centering
    \includegraphics[width=\textwidth]{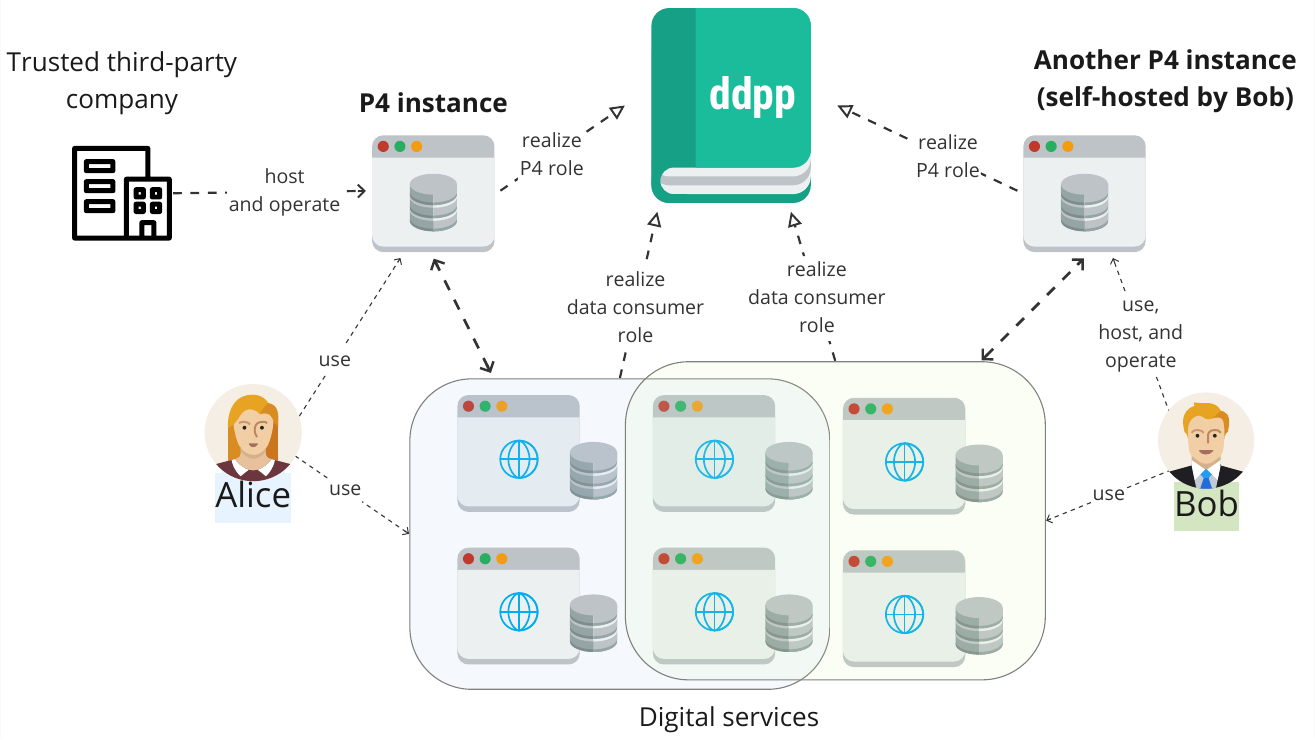}
    \caption{Overview of the decentralized data privacy protocol (ddpp).}
    \label{fig:ddpp-deployment-overview}
\end{figure}

Due to these reasons, we argue that a more adequate solution should be positioned at a higher level of abstraction and be developed as an open protocol that anyone can implement. We envision a decentralized protocol for users and services to manage data privacy preferences. A protocol can be defined as a set of syntactic and semantic rules that standardize communication between two or more entities. Using a defined protocol, users could tell the digital services \textit{where} they (the users) want to store and manage their privacy preferences. We call this place the user's \textit{Personal Privacy Preferences Place (P4)}. Users can decide who will implement their P4, where, and how. P4 instances could be hosted by a user's trusted third-party company or even be self-hosted, for example. Figure~\ref{fig:ddpp-deployment-overview} illustrates an example scenario where two users, Alice and Bob, use different sets of digital services. Alice manages her privacy preferences through a P4 instance hosted and operated by a trusted company, while Bob manages his using a self-hosted P4 instance. The digital services can interoperate with both P4 instances because they (the digital services and the P4 instances) implement their roles in the protocol.



\paragraph{Shaping the requirements}
The goal of the protocol is to improve the usability and self-sovereignty of privacy preferences management from the point of view of end users in the context of numerous digital services. To achieve this goal, the protocol must fulfill certain quality and functional requirements. From a quality perspective, the list includes (but is not limited to) \textit{openness} (the protocol shall have an open specification in order to enable operational independence, meaning that anyone could implement its elements), \textit{adaptability} (the implementation of the protocol shall not prevent service providers from specifying the privacy preferences as they want, meaning that service providers should not be obliged to change the way they define privacy preferences), and \textit{confidentiality} (the personal privacy preferences place shall not store nor exchange private user data, but only users' privacy preferences data). Concerning the functional aspects, we highlight three fundamental constructs: the \textit{data structure} (the protocol shall specify a privacy preferences meta-model whose instances express privacy preferences using generic elements), the \textit{behavior} (the protocol shall specify the interaction flows between the participants, namely the user, their P4, and the digital service), and the \textit{interfaces} (syntactic and semantic description of each interface that the participating systems -- P4 instances and digital services -- must implement to enable the desired behavior).

\paragraph{On the interaction flows} 

The protocol must support at least two key flows: \textit{handshake} and \textit{update}. Using the handshake flow, the user informs their digital service about their P4 instance. After the user sets their initial privacy preferences and after the authorization process, the digital service can communicate with the user's P4 instance to exchange privacy preferences data. Using the update flow, every change that the user makes in the privacy preferences on their P4 should be reflected in the affected digital services, and every change made in a digital service should be reflected in the user's P4.











\section{Benefits and opportunities}\label{asp24:sec:opportunities}


For users, the envisioned approach is an additional step towards enabling data sovereignty as the implementation of the protocol takes the exclusive control of privacy management means from the service providers and gives it to the actual data subjects, i.e., the users. From the point of view of the service provider, and given that we are already living in a regulated society concerning data privacy, non-compliance poses a significant financial risk. Therefore, adherence to an open standard would help service providers transfer at least part of the risk beyond the boundaries of their companies. Furthermore, adopting an open privacy management standard would increase transparency and help build trust in the service providers who adopt the protocol.

This idea can be boosted by the adoption of self-sovereign identities (SSI), which have gained increasing traction through initiatives such as the eIDAS regulation \cite{eu2023eidas}. The location of the user’s privacy preferences could be tied to their identities. When a user provides their identity to an arbitrary digital service, the service would directly know the user's P4 instance and, based on the protocol, privacy preferences would be stored on the user’s P4.


Customized and optimized P4 instances can give users extended privacy management capabilities in comparison to those offered by their digital services. For example, while a certain digital service may only allow for either consenting or denying access to certain data for a given purpose, P4 instances can provide users with the ability to set dynamic rules or constraints on their preferences (e.g., revoke consent for data \textit{X} for the purpose \textit{Y} in all digital services located in \textit{Z} 30 days after consent was granted). The development, customization, and operation of P4 instances opens up new business opportunities for companies.

Furthermore, following the same specification language and a standard ontology, we can facilitate the matching process between privacy policies and user's preferences and further automate it.


\section{Conclusion and outlook}\label{asp24:sec:conclusion}


So far, it has been hard for users to maintain sovereignty over their privacy preferences because privacy preferences management is not centered on the users but scattered across numerous digital services. In this paper, we sketched our vision of a decentralized data privacy protocol. The implementation of this protocol will allow users to manage their privacy preferences effortlessly. It primarily enables communication between web services and P4 instances but is also available for any digital service to implement and utilize.
Research plays a key role in the fulfillment of this vision. From our point of view, the next steps include a review of extensible data privacy meta-models, the design of the protocol flow, the design of a reference architecture for P4, and prototyping of a reference implementation.


%
%
%
\bibliographystyle{splncs04}
\bibliography{paper.bib}

\end{document}